\documentstyle[manuscript,osa]{revtex}

\begin{document}
\title{Analyzing Website Choice Using Clickstream Data}
\author{Avi Goldfarb\thanks{%
This research was supported by the Social Science Research Council through
the predissertation fellowship of the Program in Applied Economics and by a
Plurimus Corporation Research Fellowship. I would like to thank Plurimus
Corporation for providing me with the clickstream data and J. Walter
Thompson Company for providing me with advertising data. I would also like
to thank Shane Greenstein, Charles Manski, and Robert Porter for helpful
comments. \ All remaining errors are my own. \ Correspondence to:
a-goldfarb@northwestern.edu. 847-864-9362}}
\address{Northwestern University}
\maketitle

\begin{abstract}
This paper estimates demand for Internet portals using a clickstream data
panel of 2654 users. \ It shows that familiar econometric methodologies used
to study grocery store scanner data can be applied to analyze
advertising-supported Internet markets using clickstream data. \ In
particular, it applies the methodology of Guadagni and Little (1983) to
better understand households' Internet portal choices. \ The methodology has
reasonable out of sample predictive power and can be used to simulate
changes in company strategy. \ (JEL classification numbers: M31, C25)
\end{abstract}

\section{Introduction}

The growth of the Internet has provided economists, marketers, and
statisticians with a potentially rich and informative data source. \ Since
everything on the Internet is necessarily digital, all activity can be
easily recorded and stored in a database for future examination. \ This data
has found disparate uses, from advertisement targeting to law enforcement. \
One prevalent but relatively under used example of such data is clickstream
data. \ This data consists of each website visited by a panel of users and
the order in which they arrive at the sites. \ It is often accompanied by
the time of arrival at and departure from the site as well as the degree of
activity at the site and the demographic characteristics of the users. \
Examples of companies that collect clickstream data based on broad panels
are Netratings Inc., MediaMetrix Inc., and Plurimus Corp. \ This paper uses
data from Plurimus Corp. to analyze user choice of Internet portals. \ It
will show that commonly used econometric models for examining grocery
scanner data can be applied to clickstream data on advertising-based online
markets.

A portal is a launching pad to the Internet. \ Portals, such as Yahoo,
Lycos, and MSN, are sometimes referred to as search engines. \ Adar and
Huberman (1999 p. 2) describe them as ``a refinement of the web search
engine service''. \ Portals have search engine capabilities, but they also
have other features. \ These may include email, news, and a link-based
directory to the web separate from the search service. \ There are few, if
any, pure search engines remaining. \ In this paper, I am interested in the
portal as a starting point and not as a destination. \ Therefore I look at
the use of portal main pages, directory pages, and search pages, but not at
email, news, and shopping pages.

The methodology used here closely mimics that of Guadagni and Little's
(1983) paper that estimates a multinomial logit model with scanner data to
examine consumer coffee purchases. \ It shows that the model has reasonably
good out-of-sample predictive ability. \ Furthermore, informative
simulations can be conducted on the effects on market share of changing a
variable. \ For example, it can derive an estimate of the impact on number
of visits of increasing advertising by one dollar. \ The results, however,
have to be interpreted with caution. \ The data does not satisfy the
Independent of Irrelevant Alternatives assumption made in the model. \ This
assumption implies that there is no correlation between the alternatives
outside of the observed variables. \ When this assumption fails, estimating
switching behavior becomes difficult. \ The coefficients and the simulation
results will therefore have some bias. \ Future work will apply the
techniques of more recent developments in the econometric analysis of panel
data to the Internet portal market. \ This will alleviate the above problem.
\ 

Guadagni and Little take advantage of the richness of their data set by
treating each purchase as a separate observation. \ Those few studies that
have used clickstream data thus far (such as Goldfarb (2000a), Moe and Fader
(2000), Sandvig (2000)), have aggregated the data to a market share level. \
While this has provided interesting insights into specific problems, it is
not the best approach to understanding website choice and the causes of
website shares within a given market. \ Aggregating the data deletes
considerable relevant information. \ Important determinants of website
choice include an individual's past experience at a site and the site that
the individual went to the previous time. \ Unlike most other marketing
studies using choice-specific data, there is no monetary price here. \
Goettler and Shachar (1999) also examine a consumer panel that faces no
price consisting of individual choices of television shows.

Developing a framework to study consumer choices of free
(advertising-supported) websites is an essential step to better
understanding user behavior on the Internet. \ According to the data set
used in this study, more than two-thirds of all consumer Internet traffic is
at advertising-supported sites. \ With the exceptions of Amazon and EBay,
the top twenty sites in terms of unique visitors are all
advertising-supported. \ The literature on this important aspect of the
Internet is sparse. \ Three studies that focus on advertising-supported
websites are Adar and Huberman (1999), Gandal (2001), and Goldfarb (2000a).
\ Adar and Huberman (1999) show that portals can discriminate between users
as those looking for certain topics are willing to spend more time. \ This
means that search engines can capture more consumer surplus (in the form of
advertising revenue), but forcing consumers that are willing to spend more
time to view more pages and advertisements. \ Gandal (2001) examines market
share at an aggregate level to try to examine the portal market. \ He finds
that early entrants have an advantage and that certain features matter more
than others. \ Goldfarb (2000a) examines concentration levels in
advertising-supported Internet markets.

Lynch and Ariely (2000) is one of few Internet studies that looks at
choice-specific data. \ They construct a simulated environment for the
purchase of wine and examine purchase choice. \ Like Lynch and Ariely's
study, this paper takes advantage of the choice-specific data. \ Unlike
their study, I look at the choice of free web sites using actual user
clickstreams.

The main data for this study was supplied by Plurimus Corporation. \ It is a
clickstream data set consisting of every website visited by 2654 users from
December 27 1999 to March 31 2000. \ It also contains data on the time of
arrival at and departure from each site. \ In total, the data set contains
3,228,595 website visits, of which 859,587 (2622 people) are to Internet
portals. \ Using this data, I construct measures of past search success,
past time spent searching, whether a site is an individual's starting page,
whether an individual has an email account at the site, and the number of
pages viewed at each site. \ A considerable section of this paper is
dedicated to explaining the construction of these variables from the raw
data. \ Many of the decisions were based on a questionnaire conducted in
June 2000 (see Goldfarb 2000b for further details). I link the Plurimus data
to monthly advertising spending data from J. Walter Thompson Company and
media mentions data found through the Lexis-Nexis Academic Universe.

The next section of the paper will describe the application of the
methodology used by Guadagni and Little to the present problem. \ Section
three will describe the data set, the questionnaire used to inform data
construction, the actual process of data construction, and summary
statistics. \ Section four will present the results, test the model's
predictive ability, and examine market response to changes in the control
variables. \ The paper will conclude by summarizing the key results and
proposing several potential areas for future research.

\section{Using the Multinomial Logit With Clickstream Data}

Internet users choose which website to visit just as they make several other
economic choices: \ given the alternatives available and the information
they have about those alternatives, they choose the alternative that will
give them the highest utility. \ In terms of grocery products such as coffee
(studied by Guadagni and Little), this means that households buy the product
that has the best attributes for the lowest price. \ In terms of portals,
this means that households will use the portal that will allow them to
maximize the probability of finding what they seek and minimizing the time
spent. \ Conceptually, I assume households are exogenously given a ``goal''
when they go online. \ I explore this assumption in the questionnaire part
of this paper and in Goldfarb (2000b). \ They go to the portal that they
expect will help them achieve that goal in the least time with the most
accuracy.

In the multinomial logit model, the expected utility of the portal is based
on past history, several website characteristics (that may vary over time),
outside influences such as advertising and media mentions, and an
idiosyncratic error term. \ Formally, household $i$ visits website $j$ at
choice occasion $t$ when 
\begin{equation}
Eu_{ijt}\geq Eu_{ikt}
\end{equation}
for all $k\neq j$. \ Here $u_{ijt}$ is defined by 
\begin{equation}
Eu_{ijt}=X_{ijt}\beta _{ijt}+\varepsilon _{ijt}
\end{equation}
$X_{ijt}$ may include variables that change over any or all of $i$, $j$, and 
$t$. \ $\beta $ may vary over $i$, $j$, or $t$, implying household
heterogeneity, brand heterogeneity, time (choice occasion) heterogeneity or
any combination of the three. \ In this study, $X_{ijt}$ will never vary
over just $t$, just $i,$ just $t$ and $i,$ or just $t$ and $j$. \ It will
vary over just $j$ in the form of portal-specific dummy variables. \ $\beta $
will be assumed constant. \ Future work will look at heterogeneity across
households in $\beta $. \ There are $I$ households, $J$ websites, $T_{i}$
choice occasions for each household, and $\sum_{i=1}^{I}T_{i}$ total choice
occasions.

It is expected utility to the user, not to the observer, that is of
interest. \ It is assumed that the user knows $\varepsilon _{ijt}$. \ The
expectation is taken over relevant variables that the user may not know the
value of before visiting the website. \ For example, the user does not know
how long she will spend on the website. \ She does, however, have an
expectation of how long it will take based on her past experience at that
website.

In order to get the multinomial logit form, the $\varepsilon _{ijt}$ are
assumed to be independently distributed random variables with a type II
extreme value distribution. \ Given the above assumptions, the probability
of household $i$ choosing brand $j$ at choice occasion $t$ can be expressed
as: 
\begin{equation}
P_{it}(j|X_{ijt},\beta _{ij})=\frac{\exp (X_{ijt}\beta _{ijt})}{%
\sum_{k=1}^{J}\exp (X_{ikt}\beta _{ikt})}
\end{equation}
The model, as expressed above is a combination of Theil's (1969) multinomial
logit and McFadden's (1974) conditional logit. \ It is commonly referred to
as a mixed logit or as a multinomial logit. \ Since this paper assumes $%
\beta $ is fixed, the model here is a conditional logit. \ The log
likelihood function is as follows: 
\begin{equation}
\sum_{i=1}^{I}\sum_{t=1}^{T_{i}}\sum_{j=1}^{J}d_{ijt}\ln
P_{it}(j|X_{ijt},\beta )
\end{equation}
where $d_{ijt}$ is equal to one if alternative $j$ is chosen by individual $%
i $ at time $t$, and is equal to zero otherwise.

A significant potential problem with this framework is that it implies an
assumption of independence of irrelevant alternatives (IIA). \ If a
household is offered a new alternative that is almost identical to one of
the current alternatives, say $k$, then this new alternative should be
expected to only draw buyers from $k$; however, under IIA, the new
alternative draws buyers from all the other alternatives. \ I test for and
reject the assumption of IIA in section 4. \ This is a significant problem
that will be addressed in future work by allowing for household
heterogeneity.

In this model, the researcher observes the choice by each household on each
choice occasion. \ Let $y_{ijt}=1$ if household $i$ chooses website $j$ on
choice occasion $t$ and let $y_{ijt}=0$ otherwise. \ The researcher also
observes the characteristics of each website at that choice occasion for
that household $X_{ijt}$.

\section{Data}

\subsection{Raw data sources and description}

The main data set consists of 3,228,595 website visits by 2654 households
from December 27, 1999 to March 31 2000. \ Also included in the initial data
set was the time of arrival at and departure from a website, the beginning
and end of each online session, and the number of pages visited at that
site. \ This data, collected by Plurimus Corporation, was used to construct
a data set of 859,587 portal choices by 2622 households. This study uses
only 2008 of these households and keeps the others to test the model out of
sample. \ Furthermore, it only looks at the eight most frequently used
portals comprising eighty percent of all portal visits. \ Therefore the
final data set consists of 519,705 portal choices by 2005 households.

Plurimus has an anonymizing technology that allows them to collect
information about users without needing the users' permission. Plurimus
avoids significant privacy concerns because the users are anonymous and the
data cannot be traced to any actual person. They are regularly audited by
PriceWaterhouseCoopers in order to ensure they exceed the privacy
requirements of the FCC guidelines. \ Unlike volunteer panel data,
behavioral records from anonymized users are not biased by the wish to be
seen in a socially desirable light. \ Moreover, there is no selection bias
into the sample itself, yielding a sample from a broader spectrum of
socioeconomic status than is typically available from panel studies.

This data, however, has five limitations that need to be considered when
extending the results of this study to the entire Internet. First, the
geographic distribution of the sample is considerably biased. New York,
Chicago, and Los Angeles are under-represented. Roughly half the sample
comes from the Pittsburgh area. \ Another quarter is from North Carolina and
another eighth is from Tampa. This problem is not as severe as it may first
appear because portals are a national product.\footnote{%
Future research with Plurimus' data will not suffer from this limitation}

The second limitation is that it does not collect data on America Online
(AOL) users. \ Since AOL subscribers make up roughly 50\% of all American
home Internet users, this could bias the results. \ AOL, however, provides a
different product from the other Internet service providers. \ AOL users are
encouraged to stay within the gated AOL community and they generally do not
venture out onto the rest of the Internet. Moreover, preliminary surveys
commissioned by Plurimus show that when AOL users do leave the gated AOL
community, they have similar habits to other web users. \ This data
limitation will, however, put a downward bias on visits to the AOL portal.

Third, the data contains information on few users at work. \ Online habits
at work are likely different from those at home; however according to a
study by Nie and Erbring (2000), 64.3\% of Internet users use the Internet
primarily at home; just 16.8\% use it primarily at work. \ Few data sets,
however, contain reliable at work panel data.

Table 1 compares unique visitors as a fraction of Yahoo's users for the
eight portals used in this study as estimated by several companies. \ I
chose to use a base of comparison because the numbers vary as a result of
the assumed online population. \ I use unique visitors rather than total
visits because that was the data that was available from the other
companies. \ The number of unique visitors for a month to a website is the
number of different households that go to a given website over the course of
the month. \ Some of the variation between the methodologies may be a result
of exactly which webpages are considered part of the main site. \ The data
in the table is website-specific (not Internet property-based) meaning, for
example, that YahooSports is not considered to be a part of Yahoo. \ I could
not find website based results for Media Metrix in March or for
Nielsen/Netratings in any month. \ With the exception of AOL, Plurimus's
numbers are well within the range of the other companies, and therefore the
above issues with the data may not be important for understanding portal
choice by users who are not AOL subscribers.

The fourth limitation is that the data is collected at the household level
rather than at the individual level. \ If two people in a given household
have considerably different habits this will show up as one person with
widely varying habits. \ While this makes it difficult to assess the extent
of learning over time, it is a standard problem in consumer panels.

Fifth, it does not contain information on households from the first time
they go online. \ Therefore initial conditions are potentially a problem. \
Although the observations may not be independently and identically
distributed, this problem may be partially alleviated by the law of large
numbers due to the number of observations per household in the data set. \
More than 79\% of the households in the final data set make 30 or more
choices. \ The mean household makes 259 portal choices and the median
household makes 120 portal choices.

Together, these five data limitations mean that results should be extended
to different geographic distributions, AOL users, and at work users with
caution. \ Furthermore, the fourth and fifth limitations mean that
understanding learning behavior is not possible.

I join this clickstream data set with two other data sets. \ The first is an
advertising data set provided by J. Walter Thompson Company. \ This data set
consists of all advertising spending by each of the portals used in this
study on a monthly basis. \ The spending is determined by a thorough
sampling of television, radio, newspaper, magazine, outdoor, and Internet
advertising by each of the portals. \ The number of advertisements is then
multiplied by the average cost of advertising in each medium (at the program
level in television and the issue level in magazines). \ Since this data is
not individual-specific, it will likely underestimate the impact of
advertising. \ The methodology used in this paper, however, can easily be
adapted to individual-specific advertising data.

I also constructed a data set of `media mentions' for each of the relevant
companies. \ If a company is mentioned on network television news (ABC, CBS,
or NBC), in the Wall Street Journal, in the New York Times, or in USA Today
on a given day or the day before then the media mentions variable is equal
to one. \ Otherwise it is equal to zero. \ Unfortunately, I do not know
which individuals were actually watching or reading which media. \ It is
likely, however, that mentions in these media are highly correlated with
mentions in other media such as local newspapers.

In the data set several dozen portals are observed to be chosen. \ For
computational feasibility, I limit the number of portals to the eight with
the most visits (in order): Yahoo, Microsoft Network (MSN), Netscape,
Excite, AOL, Altavista, Iwon, and Lycos. \ These eight make up eighty
percent of all visits and all sites with more than 2.5\% of total visits. \
There was a natural break after Lycos because the ninth most visited portal,
MyWay, is a site that is the default of several Internet Service Providers
and is rarely chosen as anything but a start-up page. \ Go.com is not
included because, although it is commonly ranked in the top five portals, a
large percentage of those visits are to destination websites such as
ESPN.com, Disney.com, and MrShowbiz.com. \ The Go.com portal page itself
ranks tenth in total visits and ninth in unique visits. \ Qualitative
results, however, do not change with the addition of more portals. \ Future
work will explore methodologies that allow for the inclusion of a larger
number of portals.

\subsection{Questionnaire}

There were several issues related to analyzing a clickstream data set that
did not have obvious answers. \ I conducted an email-based survey of
Internet search habits to help resolve these issues. \ Using surveys to
inform data interpretation is relatively rare in economics. \ Helper (2000)
asserts that economists should use more surveys and field research in order
to better understand data. \ She emphasizes that this type of research
``allows exploration of areas with little preexisting data or theory'' (p.
228). \ Analysis of clickstream data certainly qualifies as one such area. \
Manski (2000) recommends questionnaires to elicit agents' preferences and
expectations directly. \ Jaffe, Trajtenberg, and Fogarty (2000) use surveys
to determine whether patent citations are a good proxy variable for
communication. \ In other words they use a survey to determine how to
interpret a data set. \ In this paper, I use a survey to determine how
derive variables such as search success from raw clickstream data. \ Further
details on the questionnaire are in Goldfarb (2000b)

\subsubsection{Questionnaire methodology}

The survey was sent to each participant as an email attachment in Microsoft
Word template format. \ In the accompanying email, I explained that I was a
doctoral student in economics studying Internet habits. \ Respondents came
from two groups. \ The first group, henceforth referred to as the `spammed'
group, consists of the 34 respondents to unsolicited email. \ The second
group of respondents consisted of 23 `friends of friends'. \ After receiving
a response rate of roughly three percent, I decided to augment my numbers by
asking several friends and family members to forward the survey to their
mailboxes. \ When there is sufficient data, I present results in this paper
for the 34 `spammed' respondents and for the 57 in the total sample.

Clearly, this is a biased sample and cannot be used to conduct classical
statistics. \ It can, however, be used to inform myths, suggest ideas, and
suggest stories. \ The survey results are quite informative about individual
surfing habits. \ By observing a biased sample of people, I can follow the
search process more closely than I can with a broader sample. \ It is common
practice in psychology and in experimental economics to draw candidates from
undergraduate classes, and then to use this information to inform theory.

The survey itself asks respondents to search for driving directions, medical
information, an MP3, and something of their own choosing. \ Respondents then
answered several questions about the searches (see the Appendix for a copy
of the full questionnaire). \ The search tasks were chosen to be diverse and
to reflect common search activities. \ The survey also asks several
questions about user Internet habits.

\subsubsection{Questionnaire Results}

Two of the issues addressed in the questionnaire are particularly important
to this paper. \ The first is determining which variables are relevant to an
analysis of search engine competition (and hence which variables to
construct and include in the study). \ The second is how to determine
whether a given search fails. \ Other issues addressed include whether
faster search is more desirable, and whether habits differ at the second
search engine in a given search from those at the first.

There are many potential relevant variables for analysis. \ The survey asked
which pages individuals bookmarked and what was each individual's starting
page. \ Individuals rarely bookmarked portals, and those that were
bookmarked were rarely used in the actual search part of the questionnaire.
\ On the other hand, start pages were found to play an important role in
site choice. \ The survey also asked respondents to give reasons why they
preferred their favorite portal. \ The only specific portal feature
mentioned was email. \ Other features such as shopping, Internet radio,
games, and an online community were not mentioned. \ As a result of these
findings, I include whether a portal is an individual's start page and
whether that person has an email account at that site. \ I do not include
other features or bookmarks.

Another important variable that the survey suggests should be included is
the goal of search. \ Most respondents claimed to use more than one search
engine because ``Different search engines are better suited to different
tasks''. \ Links to relevant pages were also said to be important. \ I could
derive data on whether a portal is linked to the next page visited. \ I did
not include goal of search in the final analysis because including it did
not satisfy the Akaike information criterion or the Bayesian information
criterion for goodness of fit.

Whether a search fails is an important factor for an individual's experience
with a data set. \ Ideally each person would only conduct one task during
each online session. \ Therefore if the researcher observes the individual
visit a search engine followed by a visit to a destination website without
searching again, then it would be reasonable to assume the search was
successful. \ In this scenario, if the researcher observes the individual
search again after going to the site then the search would appear to have
been a failure. \ More than 45\% of the respondents claim to either perform
several tasks or have no specific task in mind when they go online,
considerably complicating the definition of a failed search.

The group with no specific task in mind makes up only five percent of
respondents (6\% of spammed). \ Defining how they search and the reasons for
it are beyond the scope of this survey. \ Much more important is controlling
for the more than forty percent of respondents (also roughly 40\% of
spammed) who do several tasks when they go online. \ One way to do this is
to compare the goals of searches that occur during a given session. \ If the
goals are the same, it is more likely that they are part of the same search
task. \ Also, the elapsed time between searches may be relevant as would the
number of sites seen between the visits to search engines.

Thus, if people search twice for the same thing in a short period of time,
it seems reasonable to assume that the first search was a failure and the
second a success. \ This relies on one further assumption: that people do
not go to the destination site from the portal by typing in the name of the
site. \ They only use links on the search page. \ They may type in a
destination site, but not from the portal. \ Only 5.8\% of 155 searches
(4.7\% of 85 for spammed) were followed by the use of a non-portal site that
was not the final destination. \ This means that using the above method,
over ninety-four percent of searches labelled as successful would in fact
have been successful. \ While this is not perfect, it seems to be a
reasonable measure. \ Also, if a person goes directly from one search engine
to another then the visit to the first site is likely a failure. \ Using the
above criteria, I constructed a variable for whether each search failed. \ 

The survey also showed that more experienced users search faster. \ This
suggests that faster search is probably more desirable. \ Furthermore, the
survey suggests that habits are different at the second search engine
visited during a given search than at the first. \ Again, while the above
information comes from a statistically biased sample, it does inform the
researcher about analysis of clickstream data.

\subsection{Data set Construction}

I used the above information to construct several variables from the raw
clickstream data. \ Table 2 shows a sample of ten lines of raw data. \ Using
only this information, I constructed the following variables: email, goal of
search, start page, view length at the portal, links, search failure,
whether a portal was the first visited in the search process, and Guadagni
\& Little's weighted loyalty variable. \ I will describe the derivation of
each in turn.

A household was considered to have an email account at a site if the
household used the email feature at that site more than that at any other
portal. \ I know that a household used email at a given site because the
`host` in the data would reveal this. \ For example, `com.yahoo.mail' is
Yahoo's email provider and `com.hotmail' is MSN's email provider. \ No
household used more than one email account a large number of times, so I did
not allow for households to have more than one portal as an email provider.
\ Many households did not use a portal email provider. \ This $same\,email$
variable is potentially endogenous when individual heterogeneity is not
taken into account because users will set up an email account at their
favorite portal. \ As such it can be used as a proxy for some individual
heterogeneity. \ Furthermore, if the goal is to predict future choices or to
simulate changes, then this endogeneity is not relevant. \ It was the
initial decision to use the email that was endogenous, once that account is
set up, then each choice of portal is based on the existence of the email
account.

As described in section 3.2, knowing the goal of search is important for
knowing whether a search fails. \ The goal of search was determined by the
category of the site following a visit to a portal, if that next site was
visited within five minutes of the end of the portal visit. \ If the goal of
the search is another portal, then the goal of the first search is
considered to be the same as the goal of the second. \ If no site is visited
within five minutes of the end of a portal visit, then the search is
considered to have no known goal. \ 23.4\% of all searches have no known
goal. \ Most of these occur because many people return to a portal page
before logging off the Internet. \ I do not consider these to be failed
searches. \ The goals were divided into roughly one hundred overlapping
categories including news, music, email, shopping for computers, automotive
information and travel.

A portal is considered to be a household's $start\,page$ if at least 50\% of
all online sessions begin with that page. \ An online session is considered
to end if a user does not do any activity for thirty minutes. \ While
imperfect, this method determines a starting page for almost all of the
households. \ Like, $same\,email$, $start\,page$ is potentially endogenous.
\ People often change their start page to their favorite website. \ Again
like $same$ $email,$ this can proxy individual heterogeneity and the
endogeneity is not relevant if the goal is to predict future choices or to
simulate changes. \ 28\% of households have their start page at a portal. \
This is likely lower than the general population due to the lack of AOL
users.

The view length spent at a portal is the time of departure minus the time of
arrival (in seconds). \ Recall that it is time spent during {\it previous}
visits that is important for whether a household returns to that portal.

The number of pages viewed at a portal may reflect the depth of search. \
While individuals likely want to minimize time spent generally, search depth
may be an important control factor. As with view length, it is number of
pages viewed during previous visits that is important for whether a
household returns to that portal. \ This study only reports results from a
one period lag on $last$ $view\,length$ and $last\;number$ $of\;pages$. \
More complicated functions of past time spent and previous number of pages
viewed do not yield qualitatively different results.

Links were determined by visiting each portal and recording which websites
were directly linked to the main page. \ I recorded links in early April for
each of the portals. \ While it is possible that several of the links
changed, there were no relevant changes in partnerships over that time. \ If
the site that an individual visited following a portal visit was linked to a
portal, the $link$ variable takes on a value of one. \ Otherwise, it equals
zero. \ Note that the link variable can equal one even if the household did
not visit that portal. \ For example, a household could search for financial
information of Yahoo, and the search may turn up information on
MSNmoneycentral. \ The $link$ variable serves as a proxy for portal
features. \ Instead of listing whether a portal has features, this variable
proxies whether people actually use these features. \ In other words, if
people use a link, it means they are using a feature at that site, rather
than the search capabilities.

Search failure was constructed largely as described in section 3.2. \ If a
household visited two portal sites in a row, and there was less than five
minutes between visits, then the first search is considered a failure. \
Furthermore, if the household conducts a search and then searches again for
the same goal (at the same site or at a different one) within five minutes
of the first search then the search is considered a failure. \ While five
minutes is an arbitrary number, extending it to ten minutes or shortening it
to three minutes did not change the number of failures much. \ As with time
spent, it is whether previous searches at a site failed that matters. \ Also
as with time spent, more complicated functions of past failure do not yield
qualitatively different results. \ For robustness, I also calculated a
failed search variable that included searches that were not followed by
other searches.

If a portal was the first visited in the search process, then $%
firsttry_{ijt}=1$. \ If an individual has already searched and failed, then $%
firsttry_{ijt}=0$.

This paper mimics Guadagni and Little's methodology for constructing their
`loyalty' variable almost exactly. \ In their paper, loyalty is considered
to be a weighted average of past purchases of the brand, treated as dummy
variables. \ Let $portsame_{ijt}=1$ if household $i$ bought brand $j$ as its
previous purchase and zero otherwise. 
\begin{equation}
loyalty_{ijt}\equiv \alpha loyalty_{ijt-1}+(1-\alpha )portsame_{ijt}
\end{equation}
Rather than estimate $\alpha $ by maximum likelihood which would
significantly complicate the computational problem they calibrate $\alpha $
based on dummies for lags of length one to ten. \ In the present study, the
value for alpha that minimizes the sum of the difference between the actual
dummy coefficients and the loyalty function above was 0.7782. I also use $%
portsame$ alone as a loyalty variable in the study. \ Note that this loyalty
variable can be a result of either individual preferences for a given portal
or from some kind of lock-in. \ In future work, I plan to separate out these
effects of heterogeneity and state dependence. \ In a recent study,
Abramson, Andrews, Currim, and Jones (2000) find this to be the best loyalty
measure they tried.

In this study, I define the $portsame_{ijt}$ variable to depend on the
previous portal visited of any kind, not just the previous of the eight
portals used in this study. \ Therefore, if a household visits Yahoo then
About.com and then Yahoo again, $portsame_{ijt}$ on the second visit to
Yahoo is equal to zero, even though only two observations are included in
the data set. \ This means that a household is not considered brand loyal if
it went to a rival portal's website, even if that rival portal is not in the
sample. \ If I only included the sample, the coefficient on the loyalty
variable increases slightly but its significance falls slightly. \ Note that
the initial conditions problem frequently encountered in this literature
does not apply here due to the large number of observations per household.

How much time a household's previous visit to a portal took and whether that
search failed are only observed when the household has visited that portal
previously in the data set. \ Since not every household visits every portal,
these variables are missing for a large number of observations. \ I
therefore created a dummy variable for missing data. \ I also interact one
minus the missing data variable with the view length of previous search and
the failure of previous search variables. \ This overcomes the significant
potential bias of assuming a value for the missing data or of ignoring it
entirely. \ The missing data dummy has no economic interpretation.

Tables 3 and 4 contain descriptive statistics of the final data set.

\section{Results}

\subsection{Coefficients}

Table 5 presents the main results of the paper. \ Model (1) presents the
basic model. \ Here, the potentially endogenous variables of $same\;email$, $%
link$, and $start$ $page$ are not included. \ The variables all have the
expected signs, although $last\;view\;length$ is barely significant: $%
loyalty,$ $advertising,$ and $media$ mentions are all correlated with a
higher probability of search. \ $Last$ $view$ $length$ and $last$ $%
search\,failed$ are all correlated with a lower probability of search. \ The
positive sign on $last\,view\,length\,squared$ suggests that the effect of $%
last\,view\;length$ is concave. \ There was no expectation on the sign of $%
missing\;data.$ \ The coefficient on advertising likely underestimates the
actual effect of advertising as the data is aggregated over the month rather
than actual advertising viewed by the user.

Model (2) adds $same\;email$ and $link$ with the expected results. \ Taking
these into account makes $last\;view\;length$ significant. \ Model (3) adds $%
last$ $number$ $of$ $pages$ and $first$ $try$. \ $Last$ $number\;of$ $pages$
is found to have an increasing and concave relationship with choice
probability. \ This is consistent with the assumption that pages viewed
proxy depth of search. \ In this regression, $last$ $view\;length$ is
significant at the 99\% confidence level. \ Thus, controlling for depth,
households prefer to spend less time at a portal. \ $First\;try$ reveals
that Netscape and MSN are preferred as first pages in a search than as later
pages. \ This makes sense as they are the pages that appear when using the
search function in the Netscape Navigator and Microsoft Internet Explorer
browsers. \ They are also often default\ start pages, but the results do not
change in models (4) through (6) which control for the start page.

Model (4) adds the $start$ $page$ variable to model (2). \ The coefficient
on this variable is very large compared to the other dummy variables and the
likelihood improves more for this variable than for any others; however, the
coefficient is not significantly different from zero as it has an extremely
high standard error.

Model (5) is the same as model (4) except that is adds the interaction
variable of media mentions and loyalty. \ Of particular interest here is the
increase in the significance of $media\;mentions$. \ This suggests that
being mentioned in the media has a larger effect for households that are
less loyal to the brand.

Model (6) is the `kitchen sink' regression in that it includes all of the
variables in the study. \ The coefficients and their significance are
similar to models (1) through (5).

Another interesting aspect of all of the models is that there is a clear
brand preference for Yahoo over the others. \ Models (1) through (3) have
negative coefficients for all brand dummies (Yahoo is the base). \ Models
(4) through (6) also have negative dummies for Yahoo but others are often
preferred on the first try. \ Adding the coefficients together, however,
leaves a negative number meaning that Yahoo is prefered even on the first
try.

The Akaike information criterion revealed that $last\,view\,length\,squared$%
, $last$ $number$ $of$ $pages$ $squared,$ and $media$ $mentions\ast loyalty$
should be included. \ Other variables such as $advertising\,squared$ and $%
advertising\ast loyalty$ did not satisfy the Akaike information criterion. \
Note that including $start\;page$ increases the likelihood a great deal,
even though the effect is statistically insignificant. \ Any variables
included in this study that satisfy the Akaike information criterion also
satisfy the Bayesian information criterion.

\subsection{Robustness of coefficients to small changes}

Three different models are estimated in table 6. \ The first is the same as
model (2) except that it uses a broader definition of failed search. \ If no
search is conducted after visiting a portal, then that is included in the
failed search variable. \ Under the new definition of failed search, as
under the old definition, the coefficient is significantly negative at the
99\% confidence level; however, the magnitude of the coefficient itself is
smaller. \ Furthermore, in this regression, $last\;view\;length$ is not
significantly different from zero.

The second and third models in table 6 mimic model (2) but change the
loyalty variables. \ The second (model (8)) uses dummy variables for whether
the portal is the same as that used the previous period and that used two
periods before by that household. \ The third (model (9)) uses only the one
period lag. \ The coefficients are still significantly positive in all
cases. \ Note, however, that the explanatory power of these two methods is
considerably less than that of Guadagni and Little's loyalty variable. \ In
both of these models, the $last\;view\;length$ variable is not significantly
different from zero. \ In model (9), the coefficient becomes positive. \ If $%
last\;number\;of\;pages$ is included then $last\;view\;length$ does become
significantly negative.

Table 7 shows the results of conducting the above analysis with any seven of
the eight portals. \ Note that, with a few exceptions, the coefficients
change little. \ When either of the two largest advertisers are dropped (AOL
or Yahoo), advertising becomes insignificantly negative. When Iwon, the site
with the highest view length, is dropped, past view length becomes
insignificantly positive. \ This is not because people are playing games at
Iwon, since games are considered a destination and not part of the portal
page.

Although there is little change in the coefficients, the $\chi ^{2}$
statistics at the bottom of table 7 show that the independence of irrelevant
alternatives (IIA) assumption does not hold in this model. \ IIA implies
that there is no correlation between the alternatives outside of the effects
of known features. \ It is likely that Netscape and MSN are highly
negatively correlated since they are based on different browsers. \ This
method wrongly assumes that they are uncorrelated, bringing potential bias
to some of the coefficients and weakening the assertions that can be made
from policy analysis. \ It is essential that future work control for IIA.

These $\chi ^{2}$ statistics were calculated using a Hausman test following
Hausman and McFadden (1984). \ The coefficients on the brand dummies were
neither included in the Hausman test nor presented in table 7 although they
were estimated for each model. \ While the coefficients themselves change
little when a portal is dropped out of the estimation, the large sample size
and corresponding low variances of the coefficients lead to a rejection of
IIA. \ This is a considerable, though frequently encountered, problem\ in
this type of analysis. \ Hausman and Wise (1978) and, more recently, Berry
(1994) describe how accounting for heterogeneity alleviates this problem. \
This will be the subject of future work. \ Guadagni and Little (1983 p.
221), however, argue that ``a more important test of the model will be its
performance on a holdout sample of customers.'' \ This is conducted in the
next section.

These robustness checks suggest that the effects of $advertising$ and $%
last\;view\;length$ on probability of choice may not be significantly
different from zero. The coefficient on $start\;page\;$is also not
significantly different from zero. $\ $The effect of $media\;mentions$ is
robust, but the impact is still not large. \ The other variables are all
very important, particularly loyalty. \ The cause of the importance of the
loyalty variable, however, is unknown; it could be due to either state
dependence or unobserved household heterogeneity or both.

\subsection{Predictive Ability}

This section explores out of sample predictive power. \ Figure 1 shows the
predicted and actual shares of MSN over the fourteen weeks from December 27
1999 to March 31 2000 for an outside sample of roughly 600 households. \ The
predictions are done using both model (1) and model (6). \ In this case,
both models match the actual shares rather closely. \ Figures 2 through 8
show the predicted and actual shares for the other portals. \ The fits are
far from perfect. \ Both models under-predict Yahoo's share, both
over-predict AOL, Altavista, Excite, Iwon, and Lycos, and both fit MSN and
Netscape fairly well. \ With the exception of Iwon, model (6) fits better
than model (1). \ For each of the brands, however, both models matched the
general trends in the actual shares. Week-to-week changes in actual shares
are captured by the predicted models. \ 

Accounting for differences among households should help improve this
predictive ability. \ Preliminary work in accounting for household tastes
has shown, for example, that some people have a substantial taste preference
for Iwon, while others have a substantial dislike. \ This bimodal
distribution of tastes is averaged out in the model used here. \ Thus,
actual preferences for Iwon are not well represented. \ The preliminary work
suggests that the brands with a unimodal and narrow distribution of tastes
across households are predicted better than are other firms in the model
presented in this paper.

While not perfect, this model has significant predictive power and could be
used to explore how policies in one market would work in another.

\subsection{Market Response to Variable Changes}

Tables 8 and 9 explore the market responses to variable changes in model (2)
assuming no competitive response. \ Table 8 presents the elasticity of the
model to slight changes in the variables at the variable means. \ Table 9
converts these elasticities to changes in number of site visits. \ This
table assumes that there are a total of 76.5 million web users,
Nielsen/Netratings' estimate for the month of February, 2000. \ While the
elasticity numbers appear small, the increase in the number of site visits
from a marginal increase in a variable can be quite large. \ Taking the
results at face value, if MSN users' searches failed just 1\% less often,
MSN would get almost three million more site visits. \ If each site visit is
worth five cents (about the revenue received from the five advertisements
seen over typical two page views at a typical search engine), then it would
be worth it for MSN to implement this change as long as it cost less than
one hundred and fifty thousand dollars. \ 

The advertising results are perhaps the most interesting. \ An increase in
advertising by one dollar would bring six more visits to Altavista but
twenty-six more to Yahoo. \ Therefore, Altavista should increase its
advertising if each new site visit brings in seventeen cents of revenue and
Yahoo should increase its advertising if each new site visit brings in just
four cents of revenue.

Caution should be used in interpreting these results because of the lack of
IIA and because the functional form of the error term is important to
deriving these results. The results, however, do show what future studies
using IIA and fewer functional assumptions can achieve and they are
informative about general trends. \ For example, while the numbers
themselves may not be completely accurate, it is likely that an extra dollar
of advertising by Yahoo has a larger effect than an extra dollar of
advertising by Altavista. \ The current exercise should be viewed as an
approximation that demonstrates potential marginal gains from the variables.

Another way to simulate policy changes by the firms is to change the
underlying data and reestimate the market shares given the known
coefficients. \ This method underestimates changes because it does not count
dynamic effects. \ It does, however, provide a lower bound for the impact. \
Again using model (2), I undertook this exercise for several variables. \ If
MSN advertised as much as AOL, then MSN would gain 13,857,734 more visits
assuming 76.5 million users. \ If, on the other hand, Iwon advertised as
much as AOL then it would only gain 2,857,924 visits. \ If Lycos searches
were successful as often as Yahoo searches, Lycos traffic would rise by
25,726,505 or four percent. \ If Altavista had the same links as MSN then it
would get 98,948,093 more visitors or ten percent. \ Again, the exact
quantities of these predictions should be interpreted with caution. \ The
general trends, however, are informative.

\section{Conclusion}

This study has provided a preliminary look at estimating demand for
advertising-supported Internet websites based on clickstream data. \ The
methodology provides a reasonable fit to the actual patterns in the data. \
It has reasonable predictive power and is informative about the potential
impact of various policy changes.

This methodology has several weaknesses. \ The first, and most important, is
that it does not take into account individual heterogeneity. \ This leads to
a rejection of the Independence of Irrelevant Alternatives hypothesis as
well as poor predictive ability for Iwon and Yahoo in particular. \ In
future work with the data, I will estimate a model that accounts for this
heterogeneity.

Another weakness in this methodology is that it does not allow for the
market to grow. \ It predicts changes in share of a given population. \ It
therefore ignores the impact of new users in a rapidly growing market and
the effect of promotion on market size. \ The assumption that new users will
have similar tastes to the current ones has some supporting evidence in that
fact that market leaders change little over time in advertising-based
Internet industries (Goldfarb 2000b), but this methodology is much better at
exploring the demand of existing users rather than that of potential new
users.

With respect policy implications, the study provides a framework for
understanding policy effects. \ The simulations in section 4.3 show the
impact of potential policy changes on market shares. \ While they do not
take into account supply side reactions or individual heterogeneity, they do
give better estimates of policy effects than currently exist. \ More
detailed policy analysis can also be explored in this framework. \ For
example, a portal could simulate a link to a commonly used site, say
americangreetings.com. \ It could then determine the effect of this link on
market share. \ The actual increase in share resulting from this change
would be no more than the simulated level. \ It may be less because it may
be that people who go to a given portal are also the kind of people who like
the links it has. \ Thus the effects of the new link may be less than
predicted. \ Because it does not account for individual heterogeneity, this
model does not provide an effective framework for examining the effects of
major industry changes such as bankruptcies, nor does it provide a way to
look at the welfare impact of improved technology. \ In future work, I will
match the heterogeneous demand model to a supply side model and estimate the
effects of industry changes on demand and welfare.

The main purpose of this study was to show that demand for free online
services can be estimated using methodologies that are common in both the
economics and the marketing literature. \ The coefficients on the variables
in the study had the expected signs and the predictive ability of the model,
though not perfect, captured the major trends. \ Furthermore, informative
simulations can be conducted about the effects on share of changing variable
values. \ Clickstream data will be an important tool in understanding online
demand. \ This study has shown that the standard econometric methods that
have previously been applied to grocery scanner data can successfully be
applied to clickstream data. \ By bringing more econometric sophistication
to this analysis, economists and marketers can gain a better understanding
of online user behavior.

\section{References}

Abramson, Charles, Rick L. Andrews, Imran S. Currim, and Morgan Jones,.
``Parameter Bias from Unobserved Effects in the Multinomial Logit Model of
Consumer Choice,'' {\it Journal of Marketing Research }37 (Nov. 2000),
410-426.

Adar, Eyton, and Bernardo Huberman. ``The Economics of Surfing,'' Working
Paper No. 42, Center for eBusiness at MIT (1999).

Berry, Steven T., ``Estimating discrete-choice models of product
differentiation,'' {\it RAND Journal of Economics} 25 (Summer 1994), 242-262.

Gandal, Neil. ``The Dynamics of Competition in the Internet Search Engine
Market,'' {\it International Journal of Industrial Organization} 19 (July
2001), forthcoming.

Goettler, Ronald, and Ron Shachar, ``Estimating Brand Characteristics and
Spatial Competition in the Network Television Industry,'' \ Working Paper
No. 1999-E27, Graduate School of Industrial Administration\ Carnegie Mellon
University (1999).

Goldfarb, Avi, ``Concentration in Advertising-Supported Online Markets: An
Empirical Approach,'' Unpublished Working Paper, Northwestern University
(2000a).

Goldfarb, Avi , ``Analyzing Internet Clickstream Data,'' Unpublished Working
Paper, Northwestern University (2000b).

Guadagni, Peter M. and John D. C. Little, ``A Logit Model of Brand Choice
Calibrated on Scanner Data,'' {\it Marketing Science} 2 (Summer 1983),
203-38.

Hausman, Jerry, and Daniel McFadden, ``Specification Tests for the
Multinomial Logit Model,'' {\it Econometrica} 52 (Sept. 1984), 1219-1240.

Hausman, Jerry A., and David A. Wise, ``A Conditional Probit Model for
Qualitative Choice: Discrete Decisions Recognizing Interdependence and
Heterogeneous Preferences,'' {\it Econometrica }46 (Mar. 1978), 403-426.

Helper, Susan, ``Economists and Field Research: ``You Can Observe a Lot Just
by Watching'', \ {\it American Economic Review Papers and Proceedings }90
(May 2000), 228-32.

Hsiao, Cheng, {\it Analysis of Panel Data }(Cambridge UK: Cambridge
University Press, 1986).

Jaffe, Adam B., Manuel Trajtenberg, and Michael S. Fogarty, ``Knowledge
Spillovers and Patent Citations: Evidence from a Survey of Inventors,'' {\it %
American Economic Review Papers and Proceedings }90 (May 2000), 215-18.

Lynch, John G. and Dan Ariely, ``Wine Online: Search Costs Affect
Competition on Price, Quality, and Distribution,'' {\it Marketing Science}
19 (Winter 2000), 83-103.

Maddala, G.S., {\it Limited-dependent and qualitative variables in
econometrics }(Cambridge, UK: Cambridge University Press, 1983).

Manski, Charles, ``Economic Analysis of Social Interactions,'' {\it Journal
of Economic Perspectives} 14 (Summer 2000), 115-36.

McFadden, Daniel, ``Conditional Logit Analysis of Qualitative Behavior.'' In
P. Zarembka (ed.), {\it Frontiers of Econometrics} (New York: The Academic
Press, Inc., 1974), 105-142.

Moe, Wendy W., and Peter S. Fader, ``Capturing Evolving Visit Behavior with
Clickstream Data,''\ Unpublished Working Paper,\ Wharton Business School
(2000).

Nie, Norman H., and Lutz Erbring, ``Internet and Society.'' Unpublished
Working Paper, Stanford Institute for the Quantitative Study of Society
(2000).

Sandvig, Christian, ``The Internet disconnect in children's policy,''
Unpublished Working Paper, Stanford University (2000).

Theil, H., ``A Multinomial Extension of the Linear Logit Model,'' {\it %
International Economic Review} 10 (Oct. 1969), 251-259.

\end{document}